\documentclass[aps,twocolumn,pra,amssymb,floatfix,superscriptaddress,10pt]{revtex4}

\usepackage{graphicx}
\usepackage{mathptmx}

\begin{document}

\title{Interface dynamics of a two-component Bose-Einstein condensate driven by an external force}

\author{D. Kobyakov}
\affiliation{Department of Physics, Ume{\aa} University, 901 87 Ume{\aa}, Sweden}

\author{V. Bychkov}
\affiliation{Department of Physics, Ume{\aa} University, 901 87 Ume{\aa}, Sweden}

\author{E. Lundh}
\affiliation{Department of Physics, Ume{\aa} University, 901 87 Ume{\aa}, Sweden}

\author{A. Bezett}
\affiliation{Department of Physics, Ume{\aa} University, 901 87 Ume{\aa}, Sweden}

\author{V. Akkerman}
\affiliation{Department of Mechanical and Aerospace Engineering, Princeton University, NJ-08544-5263 Princeton, USA}

\author{M. Marklund}
\affiliation{Department of Physics, Ume{\aa} University, 901 87 Ume{\aa}, Sweden}

\begin{abstract}
The dynamics of an interface in a two-component Bose-Einstein condensate driven by a spatially uniform time-dependent force is studied. Starting from the Gross-Pitaevskii Lagrangian, the dispersion relation for linear waves and instabilities at the interface is derived by means of a variational approach. A  number of diverse dynamical effects for different types of the driving force is demonstrated, which includes the Rayleigh-Taylor instability for a constant force, the Richtmyer-Meshkov instability for a pulse force, dynamic stabilization of the Rayleigh-Taylor instability and onset of the parametric instability for an oscillating force. Gaussian Markovian and non-Markovian stochastic forces are also considered. It is found that the Markovian stochastic force does not produce any average effect on the dynamics of the interface, while the non-Markovian force leads to exponential perturbation growth.
\end{abstract}

\maketitle

\section{Introduction}

Superfluid dynamics of interfaces between Bose-Einstein condensates (BECs) of dilute atomic gases has gained active interest, both in experimental \cite{bib-1,bib-2,bib-3,bib-4,bib-4a}, and theoretical works \cite{bib-5,bib-6,bib-7,bib-8,bib-9,bib-10,bib-11,bib-12}. These studies started with the experimental realization of a multicomponent BEC \cite{bib-1,bib-2} and the theoretical investigation of the planar interface structure \cite{bib-5,bib-6,bib-7,bib-8}. At present, a good deal of the research focuses on the hydrodynamic instabilities at the interfaces in BECs, such as the Kelvin-Helmholtz instability, the Rayleigh-Taylor (RT) instability, the ferrofluid normal-field instability, and the Richtmeyer-Meshkov (RM) instability \cite{bib-10,bib-10a,bib-10b,bib-10c,bib-11,bib-14}. A common feature of these fundamental instabilities in the classical gas dynamics is the production of vortices at the unstable interface. In that sense, dynamics of an interface between two quantum fluids has some unique features with quantized vorticity being, probably, the most specific and the most interesting among them. At the same time, there is also much similarity between the classical and quantum gas dynamics, which allows "borrowing" some classical results and applying them directly or with modifications to the quantum case, see Refs. \cite{bib-10,bib-10c,bib-11,bib-14}.  Such "borrowing" is especially typical in studies of the linear instability stages, for which the difference between the classical and quantum cases is expected to be minimal. Still, the problem of rigorous derivation of such effects in BEC systems from the basic principles of quantum theory remains open for most cases. The derivation is needed not only to justify the extrapolations from the classical to quantum gas dynamics, but also to find limits of such extrapolation as well as to indicate intrinsic quantum effects involved into the problem. As an example, we point out the study of the interface waves in a system of two BECs undertaken in Ref. \cite{bib-7}. Apart from common capillary waves anticipated from the classical gas-dynamics and related to in-phase perturbations of two BECs, Ref. \cite{bib-7} demonstrated also the possibility of quantum counter-phase perturbations, for which the local interpenetration depth of two wave functions for different BECs oscillates in time. Thus, one purpose of the present paper is to provide rigorous derivation of the dispersion relations for some fundamental quasi-hydrodynamic instabilities in a two-component BEC driven by a time-dependent force. 

As a particular experimental realization of such a force, Sasaki et al. \cite{bib-10} suggested a system of two phase-segregated, interacting BECs with different spins placed in an external magnetic field gradient. This is possible, for example, for two magnetic sublevels of the $F=1$ hyperfine state of ${}^{87}\text{Rb}$ with $|F,m_F\rangle=|1,-1\rangle$ and $|1,1\rangle$. 
Starting from the Gross-Pitaevskii (GP) action for the two component BEC system, we derive the equations of motion of small perturbations of the interface.

The other purpose of the present paper is to demonstrate the diversity of hydrodynamic effects in a system of two separated BECs under the action of a time-dependent force. Recently, within classical gas dynamics, there has been increased interest in the problems of the modified RT and RM instabilities subject to time-dependent gravity, e.g. see \cite{Dimonte2000,Betti2006,Dimonte2007,Mikaelian2009}. The configuration of time-dependent gravity was discussed in the context of specially designed laboratory experiments \cite{DimonteSchneider1996}, in the scope of the problem of inertial confined fusion \cite{Betti2006,Betti1993,Kawata1993} and in relation to flame dynamics \cite{Bychkov2005,Akkerman2006,Petchenko2006,Petchenko2007}. In particular, stabilization of the RT instability by an oscillating high-frequency addition to the gravity acceleration has been proposed both in the traditional RT configuration \cite{Wolf} and for the inertial confined fusion \cite{Betti1993, Kawata1993}. Development of the parametric instability at a flame front under the action of acoustic waves and possible stabilization of the hydrodynamic flame instability has been investigated experimentally and theoretically in Refs. \cite{Searby1992,SearbyRochwerger1991,bib-19}.
Here we show that similar effects take place in BECs. Apart from the RT and RM instabilities considered before in \cite{bib-10,bib-14}, we demonstrate also the possibility of dynamical stabilization of the RT instability by high-frequency oscillations, as well as triggering of the parametric instability. Another interesting option, which has not been considered yet even in the classical gas-dynamics corresponds to a Gaussian stochastic force, which may be Markovian or non-Markovian with zero or non-zero time correlations, respectively. In the present paper we show that the Markovian stochastic force does not produce any average effect on dynamics of the interface, while the non-Markovian force leads to exponential growth of perturbations.

The paper is organized as follows. In section II we formulate the model. In section III we consider a planar (one-dimensional) quantum system with a magnetic gradient pushing the BECs towards each other, for the case of both constant and time-dependent external forces. In section IV we derive the equation, which describes the dynamics of small interface perturbations. In section V we consider interface dynamics under the action of different forces, including a constant force, a harmonic force, a pulse force, and a noisy Gaussian force. The results of the paper are summarized in section VI.

\section{The model}

The system of two BECs at zero temperature is described by two macroscopic wave functions ${\tilde \Psi _j}(\tilde t,\tilde r)$, and the mean-field GP Lagrangian \cite{bib-13}:
\begin{eqnarray}
\tilde {\cal L}(\tilde t) = \int d^3\tilde r \left[ \sum \limits_{j = 1,2}
\left\{ {{i\hbar } \over 2}\left( {\tilde \Psi _j^{\rm{*}}{{\partial {\tilde \Psi _j}} \over {\partial \tilde t}} - {\tilde \Psi _j}{{\partial \tilde \Psi _j^{\rm{*}}} \over {\partial \tilde t}}} \right) \right.\right.
\nonumber\\
+{\mu _j}|{\tilde \Psi _j}|^{2}-  \tilde V_{j}(\tilde t,\tilde r)|{\tilde \Psi _j}|^{2} - {g_{jj} \over 2}{|{\tilde \Psi _j|}^4}
\nonumber\\
\left.\left. - {{{\hbar ^2}} \over {2{m_j}}}{{\left| {\nabla {\tilde \Psi _j}} \right|}^2}\right\}- g_{12}|{\tilde \Psi _1}|^{2}|{\tilde \Psi _2}|^{2}\right],
\label{eq1}
\end{eqnarray}
where $\tilde V_{j}(\tilde t,\tilde r)$ are external potential energies of the BEC components.  The tildes in Eq. (\ref{eq1}) designate dimensional variables, which will be scaled in the following. Keeping in mind a system with two magnetic sublevels of ${}^{87}{\text{Rb}}$, we assume the same particle masses for both components ${m_1} = {m_2} \equiv m$ and the same interaction parameters ${g_{11}} = {g_{22}} \equiv g$. The interaction parameters ${g_{11}}$, ${g_{22}}$, ${g_{12}}$ are related to the respective s-wave scattering lengths ${a_{ij}}$ by ${{g_{ij}} \equiv 4\pi {\hbar ^2}{a_{ij}}/m}$. A dimensionless interaction parameter is introduced through the definition
\begin{equation}
\label{eq2}
\gamma \equiv {g_{12}}/g-1.
\end{equation}
The two components are separated in space if ${g_{12} > \sqrt {g_{11}g_{22}}}$, which implies ${\gamma  > 0}$, see \cite{bib-13}. Throughout this paper we use the condition of weak separation ${\gamma  \ll 1}$. Modification of all results of the present paper to the case of different BECs is straightforward.

The unperturbed flat interface between two components of BEC is located in the $(x,y)$ plane with $z = 0$ indicating the plane of symmetry, Fig. 1. The components $1$ and $2$ are characterized by their projection of the atom magnetic moment on the $x$-axis ${m_F} =  \pm 1$, respectively.

Using the states ${|F,m_F\rangle=|1,-1\rangle}$ and ${|1,1\rangle}$ of the atoms makes it possible to apply the Stern-Gerlach force, pushing the components in the opposite directions perpendicularly to the magnetic moment quantization axis.
We use a magnetic field whose magnitude has the gradient ${\nabla|{\textbf{B}(t)}| = \hat {\textbf{x}}d{B_x}(t)/dz \equiv \hat{\textbf{x}}B'(t)}$, which is uniform in space but may be time-dependent.
For the case of ${{}^{87}\text{Rb}}$ atoms the Stern-Gerlach potential energy in Eq.(\ref{eq1}) reads
\begin{equation}
\label{eq2a}
\tilde V_{j}(\tilde t,\tilde r)= {\tilde z} {{\mu _B B'(\tilde t)} \over {2}}(-1)^{j},
\end{equation}
where ${\mu _B}$ is the Bohr magneton, and ${1/2}$ is the Land$\acute{e}$ factor.
\begin{figure}
\includegraphics[width=3.2in,height=2.4in]{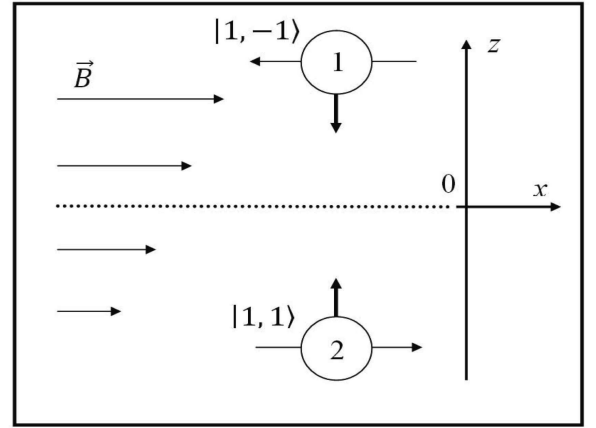}
\caption{Schematic of the two-component BEC in an external non-uniform magnetic field.}
\end{figure}

The chemical potentials $\mu _j$ in Eq.(\ref{eq1}) define normalization of stationary wave functions $\Psi _j$, and are equal to energy per particle of $j$-th condensate, which is made up of the interaction energy due to s-wave scattering and of energy of the particle in the external potential. In order to study low-energetic hydrodynamic processes in BECs we assume that energy per particle is changed negligible due to the external potential, and for normalization one can use value of the chemical potential for $\textbf{B}'=0$ in an infinite system. Such system is characterized by a uniform density sufficiently far from the interface (\textit{i.e.} on length scales much larger than the interface thickness); then the chemical potential is given by a usual result for a uniform BEC with number density ${\tilde{n}}_0$, ${\mu}=g {\tilde{n}_0}$. However, real BECs are always finite; if the length scale of our system along $z$-axis is $2 L_z$ (each condensate is of size $L_z$), then the condition that the external potential is a small perturbation, reads

\begin{equation}
\label{eq2b}
L_z {{\mu _B B'(\tilde t)} \over {2}} \ll g {\tilde{n}_0}.
\end{equation}

Dimensionless units are introduced by scaling coordinates by ${\hbar /\sqrt {m\mu } }$, time by ${\hbar /\mu }$, and atom number density (or concentration) by $\mu /g$. The dimensionless magnetic force acting on the components is
$${b}_{1}\left( t \right)=-{b}_{2}\left( t \right)=-\frac{{\hbar {\mu }_{B}}{{B}'}\left( t \right) }{2\mu \sqrt{m\mu }}\equiv -b\left( t \right).$$ The dimensionless Gross-Pitaevskii Lagrangian for two macroscopic wave functions ${\Psi _j}(t,r)$ at zero temperature reads
\begin{eqnarray}
{\cal L}(t) =  \int  {d^3}r \left[\sum \limits_{j = 1,2} \left\{
\frac{i}{2} \left(\Psi _j^* \frac{\partial \Psi _j}{\partial t} - \Psi _j
\frac{\partial \Psi _j^*}{\partial t}
 \right)  \right.\right.
 \nonumber\\
+\left| \Psi _j \right|^2- \frac{1}{2}{\left| {\nabla {\Psi _j}} \right|^2}-z{b_j}(t){\left| {{\Psi _j}} \right|^2}
\nonumber\\
-\left. \left.{1 \over 2}{\left| {{\Psi _j}} \right|^4}\right\} - (1 + \gamma ){\left| {{\Psi _1}} \right|^2}{\left| {{\Psi _2}} \right|^2} \right],
\label{eq3}
\end{eqnarray}
and equations of motion are obtained from the action principle
\begin{equation}
\label{eq4}
{{\delta S} \over {\delta \Psi _j^{\rm{*}}}} = {\partial  \over {\partial t}}\left[ {{{\delta S} \over {\delta \left( {\partial \Psi _j^{\rm{*}}/\partial t} \right)}}} \right],
\end{equation}
where the action is $S = \int {dt{\cal L}} (t)$. The result is the coupled GP equations:
\begin{eqnarray}
i{\partial {\Psi _1} \over {\partial t}} = \left[ { - {\Delta \over 2}  - 1 - z b(t)}  + {{\left| {{\Psi _1}} \right|}^2} + \left( {1 + \gamma } \right){{\left| {{\Psi _2}} \right|}^2}\right]{\Psi _1}, \\
i{\partial {\Psi _2} \over {\partial t}} = \left[ { - {\Delta \over 2}  - 1  + z b(t) + {{\left| {{\Psi _2}} \right|}^2} + \left( {1 + \gamma } \right){{\left| {{\Psi _1}} \right|}^2}} \right]{\Psi _2}.
\end{eqnarray}

\section{PLANAR DENSITY PROFILES FOR A TWO-COMPONENT BEC}
The present work is devoted to multidimensional instabilities bending an interface between two BEC components in an external magnetic field gradient. As the first step in the study, we have to specify the planar (one-dimensional, 1D) quasistationary state, which is influenced by a spatially uniform external field. In the subsections A and B we consider the cases of constant and time-dependent external fields, respectively.

\subsection{Planar profiles in the case of a stationary force}
Here we study the 1D hydrostatic state of BECs, when the constant external force ${b_0}$ is balanced by internal forces of the condensates. Let us assume that such a state was created by adiabatic switching on of the external field, and no perturbations bend the interface. The steady state is described by two real-valued functions ${\psi _{0j}}({\rm{z}})$, and the system of dimensionless GP equations reads
\begin{eqnarray}
\label{eqarray7}
0 =  {1 \over 2}{{{d^2 \psi _{01}}} \over {d{z^2}}} + {\psi _{01}} + z{b_0}{\psi _{01}} - \psi _{01}^3 - \left( {1 + \gamma } \right)\psi _{02}^2{\psi _{01}} ,\\
\label{eqarray8}
0 =  {1 \over 2}{{{d^2 \psi _{02}}} \over {d{z^2}}} + {\psi _{02}} - z{b_0}{\psi _{02}} - \psi _{02}^3 - \left( {1 + \gamma } \right)\psi _{01}^2{\psi _{02}} .
\end{eqnarray}
The time-independent force pushes the BEC components towards each other when ${b_0} > 0$. With our choice of equal masses and scattering lengths for both components, the problem possesses the symmetry ${\psi _{01}}\left( z \right) = {\psi _{02}}\left( { - z} \right)$. In the case of zero magnetic gradient the density profiles have been found in \cite{bib-7} within the limit of weak separation $\gamma  \ll 1$ as
\begin{equation}
\label{eq7}
{n_{01}}\left( z \right) = {n_{02}}\left( { - z} \right) \approx {\left[ {1 + {\rm{exp}}\left( { - 2\sqrt {2\gamma } z} \right)} \right]^{ - 1}}.
\end{equation}

We calculate the wavefunction of condensate 1 in the bulk of condensate 2. According to Eq. (\ref{eqarray7}), the wave functions penetrate into each other with the characteristic depth ${Z_p} = 1/\sqrt {2\gamma } $, see also \cite{bib-6}, which means that $\psi _{01}^2 \ll 1 $ for $z>Z_{p}$. The density profile of the bulk is described by the Thomas-Fermi  approximation (TFA), which means that $\psi _{02}^2 \approx 1-bz$ for $z>Z_{p}$. With these approximations Eq.(\ref{eqarray7}) becomes:
\begin{equation}
\label{eq8}
{1 \over 2}{{{d^2 \psi _{01}}} \over {d{z^2}}} = \left[ {\gamma  - (2+\gamma){b_0}z} \right]{\psi _{01}}.
\end{equation}
The type of solution to Eq.(\ref{eq8}) changes from exponentially decaying to oscillating in space, as $z$ becomes larger than $Z_o=\gamma/[(2+\gamma) b_0] \approx \gamma/(2 b_0)$. However, the magnitude of the spatial oscillations depends on the value $\psi_{01}(Z_o)$: it decreases exponentially with increasing $Z_o$. Therefore, the spatial oscillations become significant when $Z_o$ becomes comparable with $Z_p$, or when $b_0=\gamma \sqrt {\gamma/2}$; in this case the TFA becomes invalid, and Eqs.(\ref{eqarray7}),(\ref{eqarray8}) must be solved exactly. The spatial oscillations may be interpreted as quantum interference of matter waves, caused by interplay between the non-linearity and the dispersion (the kinetic term). This phenomenon is beyond the scope of the present work, and it will be studied in detail elsewhere. Thus, here we use parameters satisfying the condition
\begin{equation}
\label{eq10}
{b_0} \ll {b_{0cr}} \equiv \gamma \sqrt {\gamma /2},
\end{equation}
or the magnetic gradient $B^\prime$ satisfying
\begin{equation}
\label{eq9}
B^\prime < B^\prime_{0cr} \equiv \sqrt {{\gamma ^3 \mu ^3}m } /\left( {{\mu _B}\hbar } \right),
\end{equation}
which means that the spatial oscillations of the hydrostatic profiles may be neglected. It follows from Eq. (\ref{eq9}) that the critical magnetic gradient $B_{0cr}^\prime$ depends on the difference between inter- and intra-species interaction parameters as $\gamma= (g_{12} - g)/g$, and on density via the chemical potential  since  $\mu \approx g{\tilde n_0}$ in the TFA for an infinite system in a sufficiently weak external field. Therefore, the effect of quantum interference becomes stronger at lower densities of the BECs.
Equation (\ref{eq10}) yields the dimensionless condition of a well-defined interface between two condensates subject to the particular external force. In the system of two magnetic sublevels of ${}^{87}\textrm{Rb}$ we have $\gamma \simeq {10^{-2}}$ (see \cite{bib-2}), and the typical density is $5 \cdot 10^{14} {{\text{c}}{{\text{m}}^{-3}}}$. The numerical solution to Eqs. (\ref{eqarray7}), (\ref{eqarray8}) for these parameters shows that the spatial oscillations are noticeable at ${b_0} \approx {10^{ - 4}}$; so we take $b<10^{-4}$.
Numerical solutions for the wave functions near the interface are presented in Fig. 2 for ${b_0} = 7.5 \cdot {10^{ - 5}};{\text{ }}{10^{ - 4}}$, plots (a) and (b), respectively. The dashed lines in Fig. 1(a) show the approximate analytical solution of Eq.(\ref{eq7}) obtained for the case of zero magnetic field. As we can see, the analytical solution Eq.(\ref{eq7}) fits the numerical results quite well even in the case of non-zero magnetic field.
\begin{figure}
\includegraphics[width=3.6in,height=4.7in]{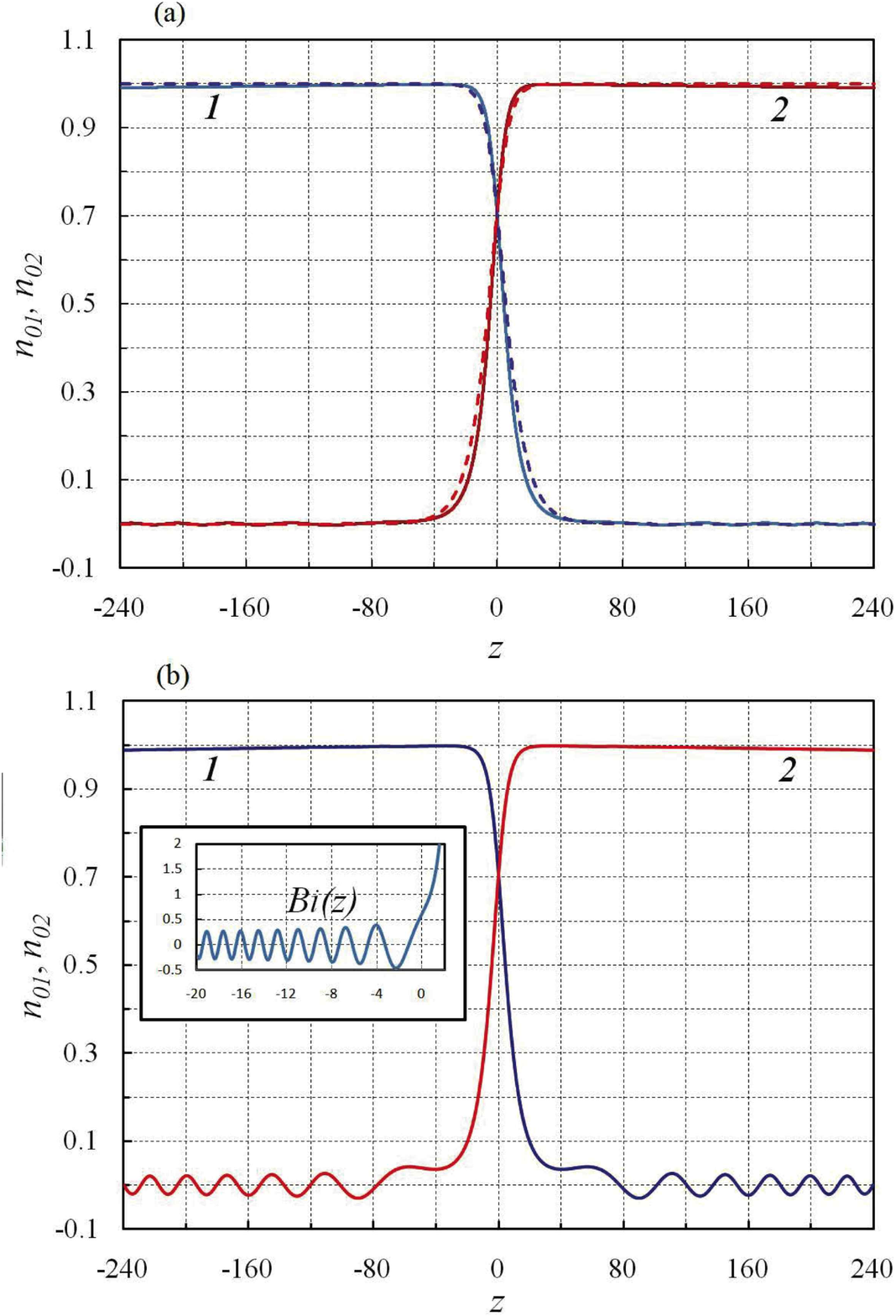}
\caption{Hydrostatic profiles of two components pushed to each other by a constant force, $b=7.5{\cdot}10^{-5}$; $10^{-4}$, figures (a) and (b), respectively. The dashed lines show the analytical solution Eq. (\ref{eq9}). The inset in figure (b) presents the Airy function of the second kind.}
\end{figure}

\subsection{1D Bogoliubov excitations of BEC due to a harmonic force}
Since we are interested in  the development of hydrodynamic instabilities for time-dependent external forces, we need to know how a harmonic force influences the planar density profiles of BECs. Similar to the constant force studied above, for time-dependent forces we also consider rather small magnetic fields, so that the respective 1D Bogoliubov excitations may be interpreted as linear acoustic waves. Another type of a time-dependent force studied in this paper is a pulse force corresponding to a weak shock in hydrodynamics. Still, it is known from hydrodynamics that a weak shock may be also described as a linear acoustic wave \cite{bib-17}.  For this reason, in the present subsection we study modifications of the BEC density in the 1D geometry due to the harmonic force.

Because of the problem symmetry with respect to $z = 0$, we consider a one-component semi-infinite BEC for $z < 0$ confined by a wall at $z = 0$. The magnitude and frequency of the force determines the energy of the generated Bogoliubov excitations. In this subsection we express density and velocity perturbations, $\delta n$ and $\delta U$, in terms of magnitude and frequency of the external force, and find the restrictions on the driving force, for which the density perturbations are small $\delta n \ll {n_0}$.

In the case of a harmonic force, the dimensionless external potential is $z{b_s}{\rm{exp}}(i{\rm{\Omega }}t)$, where ${b_s}$ is the force amplitude and ${\rm{\Omega }}$ is the frequency. Our model system is then described by a time-dependent GP equation
\begin{equation}
\label{eq11}
i{{\partial \psi } \over {\partial t}} =  - {1 \over 2}{{{\partial ^2}\psi } \over {\partial {z^2}}} + {\psi ^*}{\psi ^2} + \psi {V_{trap}}\left( z \right) + z{b_s}{\rm{exp}}(i{\rm{\Omega }}t)\psi.
\end{equation}
where ${V_{trap}}\left( z \right) =  + \infty $ for  $z < 0$ and ${V_{trap}}\left( z \right) = 0$ for  $z \ge 0$. Since the trap potential ${V_{trap}}\left( z \right)$ experiences discontinuity at the "wall", then the wave function and density are specified only for $z \ge 0$, no healing of the condensate wave function is required near the wall in TFA, and equilibrium is specified by the uniform density ${n_0} = 1$. In that case the velocity perturbation $\delta U(z,t)$ is zero at the wall $ \delta U (0,t) = 0.$ The perturbations of density and velocity,  $\delta n$ and $\delta U$, may be expressed in terms of the perturbation wave function $\delta \psi $
\begin{equation}
\label{eq12}
\psi \left( {z,t} \right) = \exp \left( { - it} \right)\left( {\sqrt {{n_0}}  + \delta \psi } \right),
\end{equation}
where $\delta \psi \left( {z,t} \right) = u\left( z \right)\exp \left( {i{\rm{\Omega }}t} \right) + {v^*}\left( z \right)\exp \left( { - i{\rm{\Omega }}t} \right)$, and $u\left( z \right)$, ${v^*}\left( z \right)$ are time-independent functions. Since the perturbations are small, we find
$$\delta n = \exp \left( {i\Omega t} \right)\left( {u + v} \right) + c.c.,$$
$$\delta U = \frac{\partial \phi}{\partial z}  = {1 \over {2i}}\exp \left( {i\Omega t} \right)
\frac{\partial }{\partial z}\left( {u - v} \right)
 - c.c.,$$
where $c.c.$ denotes complex conjugate. Then the boundary condition at the wall reads
\begin{equation}
\label{eq13}
\left. {{\partial  \over {\partial z}}\left( {u - v} \right)} \right|_{{z} = 0}^{} = 0.
\end{equation}
We linearize the GP equation (\ref{eq11}),
$$i{\partial  \over {\partial t}}\delta \psi  =  - {1 \over 2}{{{\partial ^2}} \over {\partial {z^2}}}\delta \psi  + 2\delta \psi  + \delta {\psi ^*} + z{b_s}\exp \left( {i{\rm{\Omega }}t} \right),$$
and obtain the Bogoliubov equations
\begin{equation}
- {\rm{\Omega }}u =   - {1 \over 2}{{{\partial ^2 u}} \over {\partial {z^2}}} + u + v + {b_s}z,
\end{equation}
\begin{equation}
{\rm{\Omega }}v =  - {1 \over 2}{{{\partial ^2 v}} \over {\partial {z^2}}} + v + u,
\end{equation}
which, in turn, may be reduced to
\begin{equation}
\label{eq15}
{{\rm{\Omega }}^2}\frac{\partial}{\partial z}\left( {u - v} \right) = \left[ {{{\left( { - {1 \over 2}{{{\partial ^2}} \over {\partial {z^2}}} + 1} \right)}^2} - 1} \right]{\partial  \over {\partial z}}\left( {u - v} \right) - {\rm{\Omega }}{b_s}.
\end{equation}
The solution to Eq. (\ref{eq15}) is a superposition of a running wave $ U_{1} \exp \left( { {{i \Omega }t - iqz} } \right)$  and spatially uniform oscillations $ U_{0}\exp \left( {i{\Omega }t} \right)$,
\begin{equation}
\label{eq16}
\frac{\partial}{\partial z}\left( {u - v} \right) = {\tilde U_0} + {\tilde U_1}\exp \left( { - iqz} \right),
\end{equation}
where ${\tilde U_{0,1}}$ are the respective amplitudes. The running wave is the Bogoliubov elementary excitation (the quantum acoustic wave) with the wave number $q({\rm{\Omega }})$ determined via the Bogoliubov spectrum
\begin{equation}
\label{eq16a}
\Omega ^2 = q^4/4  + q^2.
\end{equation}
It follows from the boundary condition at the wall that ${\tilde U_0} =  - {\tilde U_1} =  - {b_s}/{\rm{\Omega }}$ and
\begin{equation}
\label{eq17}
\delta U = {{{b_s}} \over {\Omega }}\left[ {\sin \left( {{\rm{\Omega }}t - qz} \right) - \sin {\rm{\Omega }}t} \right].
\end{equation}
The perturbations of density  follow from the 1D continuity equation
$$\frac{\partial }{\partial t}\delta n =  - \frac{\partial }{\partial z}\delta U$$
as
\begin{equation}
\label{eq18}
\delta n = {{{b_s}q} \over {{{\rm{\Omega }}^2}}}\sin \left( {{\rm{\Omega }}t - qz} \right).
\end{equation}
Then, the condition of weak perturbations $\delta n \ll {n_0}$ becomes ${b_s}q/{{\rm{\Omega }}^2} \ll 1$, or in the dimensional form
\begin{equation}
\label{eq18a}
{{{\mu _B}{B'}{\tilde q}} \over {2m{{\tilde \Omega} ^2}}} \ll 1,
\end{equation}
where ${\tilde q}$ and ${\tilde \Omega}$ are the dimensional wave number and frequency of the induced excitations, coupled by the Bogoliubov spectrum (\ref{eq16a}). In the low energy limit $q^4 \ll q^2$, the Boguliubov excitations are sound waves with velocity $c_s = \sqrt{g{\tilde n_0}/m}$ (unity in our dimensionless units). Then the restriction on the amplitude of the magnetic force reads 
\begin{equation}
\label{eq19}
{{{\mu _B}B'} \over {2m}} \ll {\tilde \Omega} {\sqrt {g{\tilde n}_0\over m} } .
\end{equation}
The criterion (\ref{eq19}) may be interpreted as the condition of incompressible condensates with the speed of sound much larger than any other parameter of velocity dimension involved in the problem. A more general condition (\ref{eq18a}) takes into account the Bogoliubov spectrum for higher energies. 

We solved the Bogoliubov equations for a model infinite system. The case of a finite system, confined by two walls, is solved analogously, and the boundary condition (\ref{eq13}) should be imposed for both walls. Real systems are always finite, and therefore the maximal wavelength is limited by the size of the system $L_z$. In the low-frequency limit condition (\ref{eq18a}) is equivalent to the energetic condition (\ref{eq2b}). In this limit, the Bogoliubov excitations are sound waves, with maximal wavelength of order $L_z$; substituting $L_z=2\pi {\tilde q}^{-1}$ to (\ref{eq2b}), and using the sound spectrum, one obtains (\ref{eq19}).

\section{LINEAR ANALYSIS OF THE INTERFACE DYNAMICS}
We now consider the dynamics of the interface between two infinite BECs in the presence of an external force, which may be time-dependent, $b(t)$. Similarly to Refs. \cite{bib-7,bib-13,bib-16} we use a variational approach. We represent the wave functions in terms of density ${n_j}$ and velocity potential ${\phi _j}$ as ${\psi _j}\left( {t,x,z} \right) = \sqrt {{n_j}} \exp \left( {i{\phi _j}} \right)$, so that the action for the system of two BECs is
\begin{eqnarray}
S =  \int dt \int  d^3 r \left\{ \sum \limits_{j = 1,2}  \left[ n_j
\frac{\partial \phi _j}{\partial t}
+ {1 \over 2} n_j \left( \nabla \phi_j \right)^2
   \right.\right.+
\nonumber\\
\left.  {1 \over 2} \left( \nabla \sqrt {n_j}  \right)^2 - n_j \right] + z\,b(t) \left( n_2 - n_1 \right) +
 \nonumber\\
 \left. {1 \over 2}\left(n_1^2 +n_2^2 \right) + \left( 1 + \gamma  \right)n_1 n_2
 \right\}
 .
\label{eq20}
\end{eqnarray}
For the variational calculation, we employ an ansatz that is able to describe bending and compression of the interface between the two condensates. We are interested in perturbation wavelengths much larger than the interface thickness, $kZ_p = k/\sqrt{2\gamma} \ll 1$, so that the density deformations may be treated as a bending of the unperturbed profile as a whole,
\begin{equation}
\label{eq21}
n_j(z,x,t) = n_{j0}\left[z-\delta z_j(x,t)\right],
\end{equation}
where $n_{j0}$ are the stationary density profiles for the BECs.
The variational functions are taken to be of the form of Fourier modes with separation of space and time dependence,
\begin{equation}
\delta z_j(x,t) = \zeta_j(t) \cos(k x),
\label{eq22a}
\end{equation}
\begin{equation}
\phi_j(z,x,t) = \xi_j(t) f_j(z) \cos(k x).
\label{eq22}
\end{equation}
In classical hydrodynamics, perturbations of both components of an inert interface are related by the continuity condition $\zeta_1=\zeta_2$.
In the quantum gas-dynamics of two BECs this is not necessarily the case, since we may have interpenetration of the wave functions.

The Lagrange equations with respect to the variables $\xi_j$ are
\begin{equation}
\dot{\zeta}_j \int dz\, n_j' f_j   = \xi_j \int dz \left [ f_j \frac{d}{dz}\left(n_j f_j'\right) + k^2 f_j^2n_j \right].
\label{eq23}
\end{equation}
These Lagrange equations are satisfied if the respective continuity equations are
\begin{equation}
\label{eq24}
{\dot\zeta}_j n_j'= \xi_j\left[ \frac{d}{dz}\left(n_j f_j'\right) + k^2 f_jn_j\right].
\end{equation}
First, the time-dependent terms on each side have to be equated, which, up to constant factors, leads to
\begin{equation}
\label{eq25}
\dot\zeta_j(t) = \xi_j(t).
\end{equation}
This reflects the fact that $\xi_j$ and $\zeta_j$ are collective canonically conjugate variables.
For the functions $f_j$, Ref.\cite{bib-7} gave the approximate solution
\begin{equation}
\label{eq26}
f_1(z) =-f_2(-z) \left\{\begin{array}{ll}
-(1+kz)^{-1}, & {\rm if } \quad z<0,\\
-k^{-1} \exp (-kz), & {\rm if } \quad z>0.
\end{array}
\right.
\end{equation}
The same solution holds in the presence of a small external force under the conditions specified in Sec.\ III.
We take Eq. (\ref{eq26}) as our ansatz for $f_j$. More elaborate expressions for $f_1(z)$, and $f_2(z)$ may provide a smooth transition in the interface region; however, the ansatz Eq. (\ref{eq26}) yields reliable analytical results, as we will demonstrate below by comparison to the numerical solution to the problem.

We now derive the Lagrange equation for $\zeta_j(t)$.
No time derivatives in $\zeta_j$ appear in the Lagrangian, so the Lagrange equation reads
\begin{equation}
\label{eq27}
0 = \frac{\delta S}{\delta \zeta_j} = I_{1(j)} + I_{3(j)} + I_{5(j)}+ I_{7(j)},
\end{equation}
where we introduce $I_{l(j)}$ to label the nonvanishing terms of $\delta S/\delta \zeta_j$; the numbering $l$ in $I_{l(j)}$ refer to the respective terms in Eq. (\ref{eq20}). For convenience, we assume a finite size of the system, $L_z$, $L_x$, though the final result does not depend on the size restrictions. We have
\begin{eqnarray}
&&I_1 =\frac{\delta}{\delta \zeta_j} \int {d^2}\textit{\textbf{r}}\, n_j \frac{\partial \phi_j}{\partial t}= \nonumber\\
&&\int dz\,dx\, n_{0j}'(z+\delta z_j)f_j \dot{\xi_j}\cos^2(kx)=  \nonumber\\
&&\frac{L_x}{2} \dot{\xi_j} \int dz\, n_{j}'(z) f_j(z).
\label{eq28}
\end{eqnarray}
The next nonvanishing term comes from the zero-point motion energy,
\begin{eqnarray}
&&{I_3} \equiv {\delta  \over {\delta {\zeta _j}}} \int {d^2}\textit{\textbf{r}}\quad{1 \over 2}{\left( {\nabla \sqrt {{n_j}} } \right)^2}= \nonumber\\
&&{1 \over 2}{\delta  \over {\delta {\zeta _j}}} \int {d^2}\textit{\textbf{r}}\,\left[ {{{\left( {{\partial  \over {\partial x}}\sqrt {{n_j}} } \right)}^2}+{{\left( {{\partial  \over {\partial z}}\sqrt {{n_j}} } \right)}^2}} \right]= \nonumber\\
&&{1 \over 2}{\delta  \over {\delta {\zeta _j}}} \int {d^2}\textit{\textbf{r}}\,{\left( {{\partial  \over {\partial z}}\sqrt {{n_j}} } \right)^2}\left[ {1 + {{\left( {{\partial  \over {\partial x}}\delta {z_j}} \right)}^2}} \right].
\label{eq29}
\end{eqnarray}
The first term within the parentheses yields a constant independent of $\zeta$, and the second term equals
\begin{eqnarray}
&&{I_3} = {1 \over 2}{\delta  \over {\delta {\zeta _j}}} \int  {d^2}\textit{\textbf{r}}\,{\left( {{\partial  \over {\partial z}}\sqrt {{n_j}} } \right)^2}{\left( {{\partial  \over {\partial x}}\delta {z_j}} \right)^2}=\nonumber\\
&&{{{L_x}} \over 2}{1 \over 2}{\delta  \over {\delta {\zeta _j}}} \int  dz{\left( {{d \over {dz}}\sqrt {{n_{0j}}} } \right)^2}{\left( {{\zeta _j}k} \right)^2}=\nonumber\\
&&{{{L_x}} \over 2}{\zeta _j}{k^2} \int \limits_{-\infty }^{ + \infty } dz{\left( {{d \over {dz}}\sqrt {{n_{0j}}} } \right)^2}.
\label{eq30}
\end{eqnarray}
The next non-vanishing term corresponds to the perturbation of the potential energy of the system,
\begin{eqnarray}
&&{I}_{5\left(j\right)} = {\delta\over{\delta {\zeta _j}}} \int  {d^2}\textit{\textbf{r}} \,z\, b(t)\left( {{n_2} - {n_1}} \right)=\nonumber\\
&&{{L_x} \over 2}{\left({-1}\right)}^{j} b(t)  {\delta  \over {\delta {\zeta}_{j}}}  \int  dz\,z\,{n_j}=\nonumber\\
&&{{L_x} \over 2}{\left( { - 1} \right)^j}b(t){\delta  \over {\delta {\zeta _j}}}  \int  dz\,z\,{n_{0j}}\left( {z + \delta {z_j}} \right).
\label{eq31}
\end{eqnarray}
It is more convenient to calculate these terms for $j = 1$ and for $j = 2$ separately. Using the coordinate transformation ${z'} = z + \delta {z_j}$ we modify the last expression as
\begin{eqnarray*}
&&{I_{5\left( 1 \right)}} =  - {{L_x} \over 2}b(t){\delta  \over {\delta {\zeta _1}}}\int \limits_{-{L_z}}^{ + \infty } dz\,z\,{n_{01}}\left( {z + \delta {z_1}}\right)=\\
&&- {{L_x} \over 2}b(t){\delta  \over {\delta {\zeta _1}}} \int \limits_{-{L_z} + \delta {z_1}}^{ + \infty } dz'\left[ {z'{n_{01}}\left( {z'} \right) - \delta {z_1}{n_{01}}\left( {z'} \right)} \right],
\end{eqnarray*}
\begin{eqnarray*}
&&{I_{5\left( 2 \right)}} = {{L_x} \over 2}b(t){\delta  \over {\delta {\zeta _2}}} \int \limits_{-\infty }^{{L_z}} dz\,z\,{n_{02}}\left( {z + \delta {z_2}}\right)=\\
&&{{L_x} \over 2}b(t){\delta  \over {\delta {\zeta _2}}} \int \limits_{-\infty }^{{L_z} + \delta {z_2}} dz'\left[ {z'{n_{02}}\left( {z'} \right) - \delta {z_2}{n_{02}}\left( {z'} \right)} \right].
\end{eqnarray*}
The integrals can be calculated as
\begin{equation}
\label{eq32}
{I_{5(1)}} =  -{{L_x} \over 2} b(t){\delta  \over {\delta {\zeta _1}}} \left( { - {1 \over 2}\delta z_1^2 + \delta z_1^2} \right) =  - {{{L_x}} \over 2}b(t){\zeta _1},
\end{equation}
\begin{equation}
\label{eq33}
{I_{5(2)}} = {{L_x} \over 2}b(t){\delta  \over {\delta {\zeta _2}}} \left( {{1 \over 2}\delta z_2^2 - \delta z_2^2} \right) =  - {{{L_x}} \over 2}b(t){\zeta _2}.
\end{equation}
The next term of $\delta S/\delta {\zeta _j}$ does not contribute to the equations of motion, while the last (seventh) term ${I_7}$ is important. Analysis of the terms calculated so far shows that the variables ${\zeta _j}$ are governed by two coupled oscillator equations, and the coupling happens because of ${I_7}$. Since the equations for ${\zeta _1}$ and ${\zeta _2}$ are identical, the normal modes may be in-phase, ${\zeta _1} = {\zeta _2}$, and counter-phase, ${\zeta _1} =  - {\zeta _2}$. First, let us calculate ${I_7}$ for the case of ${\zeta _1} = {\zeta _2}$, which we denote by ${I_{7(in)}}$. In this case $\delta {z_1} = \delta {z_2} \equiv \delta z$, and we find
\begin{eqnarray}
I_{7(in)} = {\delta  \over {\delta {\zeta _j}}} \int  {d^2}r\left( {1 + \gamma } \right){n_1}{n_2}=\nonumber\\
{{L_x} \over 2}{\delta  \over {\delta {\zeta _j}}} \int \limits_{-\infty }^{ + \infty } dz\left( {1 + \gamma } \right){n_{01}}\left( {z + \delta z} \right){n_{02}}\left( {z + \delta z} \right) = 0.
\label{eq34}
\end{eqnarray}
Next we consider the case of counter-phase oscillations with $\delta {z_1} =  - \delta {z_2} \equiv \delta z$ and $\zeta _1 =  - \zeta _2\equiv \zeta$, which leads to
$${I_{7(co)}} = {{L_x} \over 2}{\delta  \over {\delta \zeta }} \int \limits_{-\infty }^{ + \infty } dz'\left( {1 + \gamma } \right){n_{01}}\left( {z'} \right){n_{02}}\left( {z' - 2\delta z} \right).
$$
To calculate $\delta /\delta \zeta $ in the last expression, it is easiest to assume $\delta z$ small and Taylor expand ${n_{02}}\left( {z' - 2\delta z} \right)$ to second order, which results in
$${I_{7(co)}} = {{{L_x}} \over 2}4\left( {1 + \gamma } \right)\zeta  \int \limits_{-\infty }^{ + \infty } dz\,{n_{01}}{{{d^2}{n_{02}}} \over {d{z^2}}}.$$
The last result may be transformed to a symmetric form similar to \cite{bib-7}:
\begin{equation}
\label{eq36}
{I_{7(co)}} =  - {{{L_x}} \over 2}4\left( {1 + \gamma } \right)\zeta  \int \limits_{-\infty }^{ + \infty } dz\,{{d{n_{01}}} \over {dz}}{{d{n_{02}}} \over {dz}}.
\end{equation}
Taking into account all the terms obtained above we write equations of motion for ${\zeta _j}$. The integrals are calculated using the density profile (\ref{eq7})
\begin{equation}
\label{eq37}
 \int \limits_{-\infty }^{ + \infty } dz\,{{d{n_{0j}}} \over {dz}}{f_j} = 1/k,
\end{equation}
\begin{equation}
\label{eq38}
 \int \limits_{-\infty }^{ + \infty } dz\,{\left( {{d \over {dz}}\sqrt {{n_{0j}}} } \right)^2} = \sqrt {\gamma /8},
\end{equation}
\begin{equation}
\label{eq39}
 \int \limits_{-\infty }^{ + \infty } dz\,{{d{n_{01}}} \over {dz}}{{d{n_{02}}} \over {dz}} =  - \sqrt {8\gamma } /3.
\end{equation}
We designate the interface perturbations of the in-phase oscillations by  ${\zeta _1} = {\zeta _2} \equiv {\zeta _{in}}$ and obtain the final equation
\begin{equation}
\label{eq40}
{{{\partial ^2}{\zeta _{in}}} \over {\partial {t^2}}} + {\zeta _{in}}\left[ {{k^3}\sqrt {\gamma /8}  - b(t)k} \right] = 0.
\end{equation}
In the opposite case of the counter-phase perturbations ${\zeta _1} =  - {\zeta _2} \equiv {\zeta _{co}}$ we find
\begin{equation}
\label{eq41}
{{{\partial ^2}{\zeta _{co}}} \over {\partial {t^2}}} + {\zeta _{co}}\left[ {{k^2}\sqrt {{\gamma  \over 8}}  + {4 \over 3}\left( {1 + \gamma } \right)\sqrt {8\gamma }  - b(t)} \right]k = 0.
\end{equation}
In the limiting case of $b = 0$ our theory reproduces the results of \cite{bib-7}. Within the limit of a weak field, $b \ll {b_m} = \gamma \sqrt {2\gamma } $, and long wavelength perturbations $k/\sqrt {2\gamma }  \ll 1$, the terms in the parenthesis of Eq. (\ref{eq41}) are of different orders of magnitude, with the term $4\left( {1 + \gamma } \right)\sqrt {8\gamma } /3$ dominating. For this reason, to the leading terms, the counter-phase mode is not affected by the external potential, and we do not consider it in the present work. In the following, we work only with the in-phase mode omitting the subscript "in" as $\zeta \left( t \right) \equiv {\zeta _{in}}(t)$, and use Eq. (\ref{eq40}) as the starting point for the subsequent analysis.

\section{DYNAMICS UNDER DIFFERENT TYPES OF TIME-DEPENDENT DRIVING FORCE}
In this section we demonstrate the diversity of hydrodynamic effects in a system of two separated BECs under the action of a time-dependent force. In particular, we are interested in the RT instability developing due to a constant force, in the RM instability for a pulse force, in the parametric instability for an oscillating force, and in the interface dynamics under the action of a stochastic Gaussian force.

\subsection{The RT instability driven by a constant force}

Here we consider a constant force $b\left( t \right) = {b_0} = const$, which drives the RT instability and capillary-gravitational waves in the unstable or stable configurations, respectively.   Constant coefficients of Eq. (\ref{eq40}) lead to harmonic oscillations of the interface $\zeta \left( t \right) \propto {\rm{exp}}(i\omega t)$ with the dispersion relation
\begin{equation}
\label{eq42}
{\omega ^2} = {k^3}{{\sqrt \gamma  } \over {2\sqrt 2 }} - k{b_0},
\end{equation}
which may be written in dimensional form as
\begin{equation}
\label{eq43}
{\tilde \omega ^2} = {\tilde k^3}{{{\hbar ^2}\tilde n_0^{1/2}} \over {2{m^2}}}\sqrt {2\pi ({a_{12}} - a)}  - \tilde k{{{\mu _B}B'} \over {2m}},
\end{equation}
where $\tilde \omega$  and $\tilde k$  are the dimensional perturbation frequency and wave number. Note that Eq. (\ref{eq42}) has the same form as the dispersion relation for capillary-gravitational waves in classical hydrodynamics \cite{bib-17}. When the external magnetic field is zero, we find capillary waves with frequency related to the wave number as
\begin{equation}
\label{eq44}
{\rm{\Omega }}_{\rm{c}}^2 = {k^3}\sqrt {\gamma /8}.
\end{equation}
In the case of positive gradient of the magnetic field the dispersion relation Eq. (\ref{eq42}) yields ${\omega ^2} < 0$ for perturbations with sufficiently long wavelengths, which corresponds to the RT instability $\zeta \left( t \right) \propto {\rm{exp}}({\alpha _0}t)$ with the growth rate depending on the wavelength as
\begin{equation}
\label{eq45}
\alpha _0^2 = {b_0}k(1 - {k^2}/k_c^2)
\end{equation}
and with the cut-off wave number
\begin{equation}
\label{eq46}
{k_c} = {\left( {8/\gamma } \right)^{1/4}}\sqrt {{b_0}} .
\end{equation}
According to Eq. (\ref{eq42}), the maximal growth rate corresponds to ${k_{max}} = {k_c}/\sqrt 3 $. Figure 3 compares the analytical results of Eq. (\ref{eq42}), (\ref{eq43}) to the numerical solution to the linearized GP equations. The numerical solution is obtained by diagonalization of the linearized GP equations, approximated by finite difference method on a grid of 3335 points on the interval $\left[ { - 240;240} \right]$ in the dimensionless coordinates. Figure 3 shows good agreement of the analytical theory with the numerical results. The instability growth rate Eq. (\ref{eq45}) agrees also with the formulas employed in \cite{bib-10}.
\begin{figure}
\includegraphics[width=3.5in,height=2.5in]{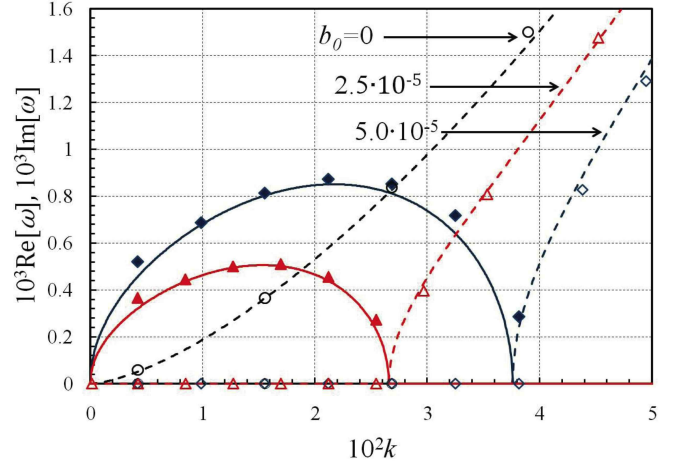}
\caption{Lowest Bogoliubov modes (the RT dispersion relation) for different values of the driving force. The analytical result is obtained from Eq. (\ref{eq42}), and presented by lines. Dashed lines show real parts of the Bogoliubov eigenvalues (waves), and solid lines correspond to the imaginary parts (instability). The markers show respective numerical solution to the linearized GP equations; filled markers correspond to the instability regime and empty markers correspond to waves.}
\end{figure}

Though the present work considers only the linear stage of the RT instability, for illustrative purposes, it is worth presenting the results obtained for the nonlinear stage. For example, Fig. 4 shows the development of the RT instability at the nonlinear stage in a confined system of two BECs obtained numerically using the methods of \cite{bib-14}. In the numerical solution, we begin with a system of $3.2 \cdot {10^7}$ atoms of ${}_{}^{87}{\rm{Rb}}$, equally split between the two states, with $\left| {F,{m_F}} \right\rangle  = \left| {1, - 1} \right\rangle $ and $\left| {1,1} \right\rangle $. We use the "pancake" trap geometry with ${\tilde \omega _x} = {\tilde \omega _z} = 2\pi  \cdot 100{\rm{Hz}}$ and ${\tilde \omega _y} = 2\pi  \cdot 5{\rm{kHz}}$. We solve the coupled GP equations numerically to find the ground state of the system. The two components occupy the regions with  $z > 0$ and $z < 0$, respectively. A perturbation of size $9{\rm{\mu m}}$ is used, and the field of view for all images is $55{\rm{\mu m}}$ square. The magnetic field of magnitude $B' = 1.78{\rm{G}}/{\rm{m}}$ is added to the system, directed so that the two components are pushed toward each other. The system is shown at times 15.9, 19.08, 22.26, 25.44 (ms) after the magnetic gradient is added. The nonlinear stage may be described qualitatively as development of the initial sinusoidal perturbations into a mushroom structure with subsequent generation of quantum vortices. More detailed investigation of the nonlinear stage of the RT instability will be presented elsewhere. Similar results may be found also in \cite{bib-10}.
\begin{figure}
\includegraphics[width=3.4in,keepaspectratio]{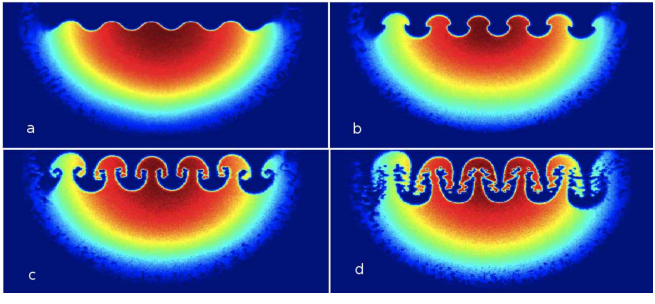}
\caption{Development of the RT instability in a trapped system of two-component BEC presented for one of the components.}
\end{figure}

\subsection{High-frequency stabilization of the RT instability due to an oscillating force and parametric instability}
In this subsection we consider an oscillating force $b\left( t \right) = {b_0} + {b_s}\cos \left( {{\rm{\Omega }}t} \right)$, where ${b_0}$ and ${b_s}$ are some constant amplitudes and ${\rm{\Omega }}$ is the frequency of the force. We show that an oscillating force leads to two interesting effects: 1) High-frequency oscillations may stabilize the RT instability produced by the constant background force ${b_0}$; 2) The oscillating force may trigger a parametric instability even in the absence of any constant acceleration, e.g. for ${b_0} = 0$.
We start with the first effect, which is similar to the mechanical phenomenon known as the Kapitsa pendulum \cite{bib-18}. We assume that the frequency ${\rm{\Omega }}$ of the external force is much larger than the RT instability growth rate ${{\rm{\alpha }}_0} = \sqrt {{b_0}k(1 - {k^2}/k_c^2)} $, that is, $\Omega \gg \alpha _0$. Then we separate the slow averaged terms (labeled "a") and fast oscillating terms (labeled "f") in the solution to Eq. (\ref{eq40}) as $\zeta  = {\zeta _a}\left( t \right) + {\zeta _f}\left( t \right)\cos \left( {{\rm{\Omega }}t} \right)$, where ${\zeta _a}\left( t \right)$ and ${\zeta _f}\left( t \right)$ change slowly on the time scale of high-frequency oscillations $2\pi /{\rm{\Omega }}$. Following the calculation method of \cite{bib-19}, we take into account the relation $\ddot \zeta  \approx {\ddot \zeta _a} - {{\rm{\Omega }}^2}{\zeta _f}\cos \left( {{\rm{\Omega }}t} \right)$ and modify Eq. (\ref{eq40}) to
$${\ddot \zeta _a} - {{\rm{\Omega }}^2}{\zeta _f}\cos \left( {{\rm{\Omega }}t} \right) = {b_0}k\left( {1 - {{{k^2}} \over {k_c^2}}} \right)\left[ {{\zeta _a} + {\zeta _f}\cos \left( {{\rm{\Omega }}t} \right)} \right]$$
$$ + {b_s}k\left[ {{\zeta _a}\cos \left( {{\rm{\Omega }}t} \right) + {1 \over 2}{\zeta _f} + {1 \over 2}{\zeta _f}\cos \left( {2{\rm{\Omega }}t} \right)} \right].$$
Separating the terms oscillating with different frequencies we find the relation between the amplitudes of the slow and fast terms
$${\zeta _{2f}} =  - {{{b_s}k} \over {{{\rm{\Omega }}^2} + {b_0}k(1 - {k^2}/k_c^2)}}{\zeta _a} \approx  - {{{b_s}k} \over {{{\rm{\Omega }}^2}}}{\zeta _a}$$
and reduce Eq. (\ref{eq40}) to
\begin{equation}
\label{eq49}
{\ddot \zeta _a} = {b_0}k\left( {1 - {{{k^2}} \over {k_c^2}} - {{b_s^2k} \over {2{b_0}{{\rm{\Omega }}^2}}}} \right){\zeta _a}.
\end{equation}
Equation (\ref{eq49}) describes stabilization of the RT instability by high-frequency oscillations for
$$1 - {{{k^2}} \over {k_c^2}} - {{b_s^2k} \over {2{b_0}{{\rm{\Omega }}^2}}} < 0.$$
Thus, stabilization of the RT perturbations of wave number $k$ by the high-frequency term is expected for the oscillation amplitude
\begin{equation}
\label{eq50}
{{{b_s}} \over {{b_0}}} > \sqrt 2 {{\rm{\Omega }} \over {\sqrt {{b_0}k} }}\sqrt {1 - {{{k^2}} \over {k_c^2}}},
\end{equation}
which may be written in dimensional units as
\begin{equation}
\label{eq51}
{{B{'_s}} \over {B{'_0}}} > \sqrt {{{4m{{\rm{\Omega }}^2}} \over {{\mu _B}B{'_0}\tilde k}}\left( {1 - {{{{\tilde k}^2}} \over {\tilde k_c^2}}} \right)}.
\end{equation}
\begin{figure}
\includegraphics[width=3.5in,height=2.0in]{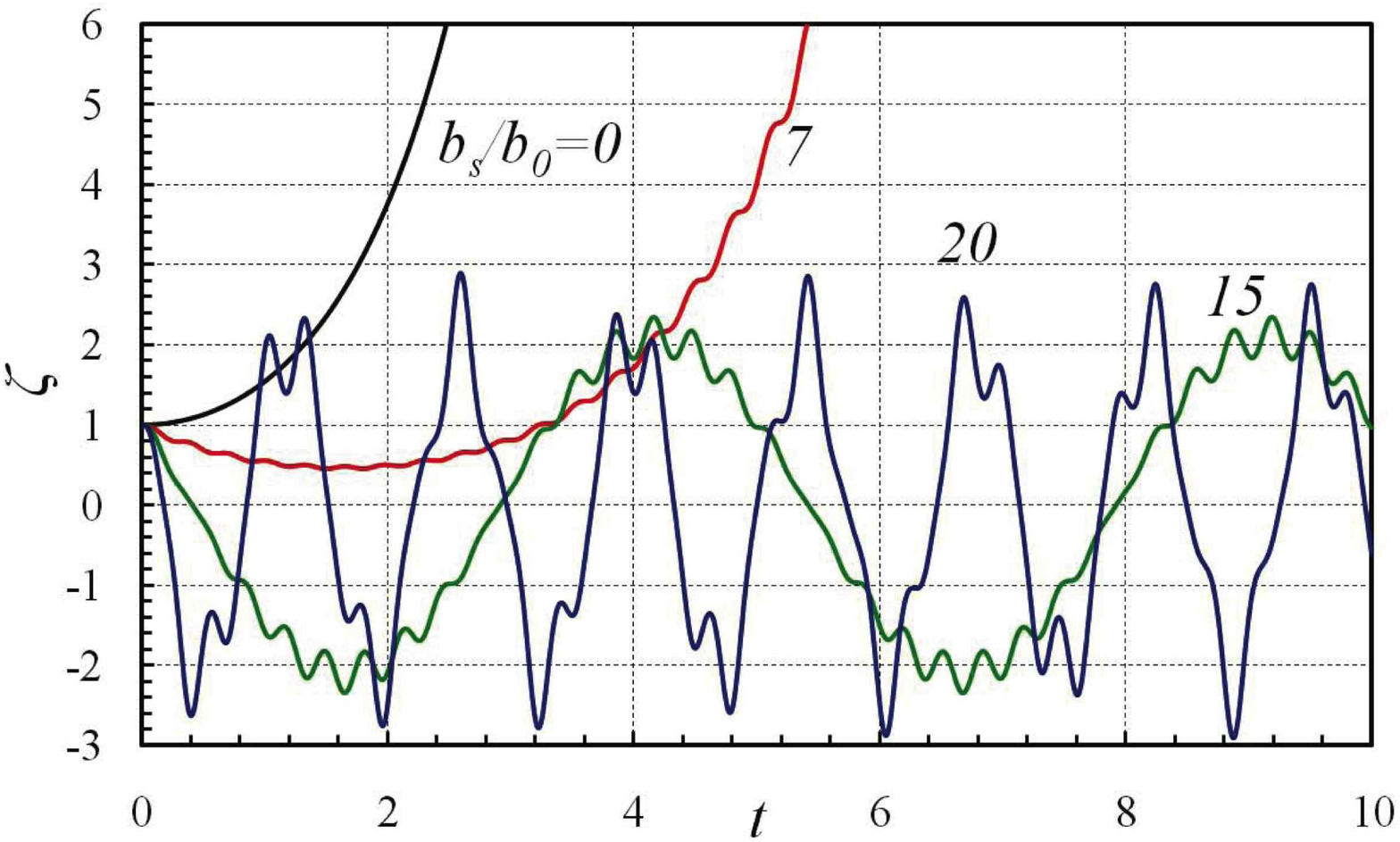}
\caption{Stabilization of the RT instability by a high-frequency force for the amplitude ratios ${b_s}/{b_0} = 0;7;15;20$, for ${b_0} = 2.5 \cdot {10^{ - 5}}$, $k = {k_{max}}$ and ${\rm{\Omega }} = 20{\alpha _0}$.}
\end{figure}
The stabilization of the RT instability by high-frequency oscillations is illustrated in Figure 5, which presents the results of direct numerical simulation of Eq. (\ref{eq40}) for the amplitude ratios ${b_s}/{b_0} = 0;7;15;20$, for the stationary component of the magnetic field ${b_0} = 2.5 \cdot {10^{ - 5}}$, the wave number corresponding to the maximal growth rate $k = {k_{max}}$ and frequency of the oscillating magnetic field  component ${\rm{\Omega }} = 20{\alpha _0}$. In agreement with the theoretical predictions, unstable exponential growth of perturbations (in average) for ${b_s}/{b_0} = 0;7$ is replaced by waves at the interface for sufficiently large amplitudes of the oscillating part of the magnetic field, ${b_s}/{b_0} = 15;20$. Figure 6 shows stability limits in the parameter space of the scaled wave number $k/{k_c}$  and the oscillation amplitude ${b_s}/{b_0}$; the shading indicates the absolute value of the instability growth rate in the unstable region. Figure 7 shows the RT instability growth rate in the presence of the external high-frequency magnetic field plotted versus the perturbation wave number for different values of the external field.
\begin{figure}
\includegraphics[width=2.9in,height=2.4in]{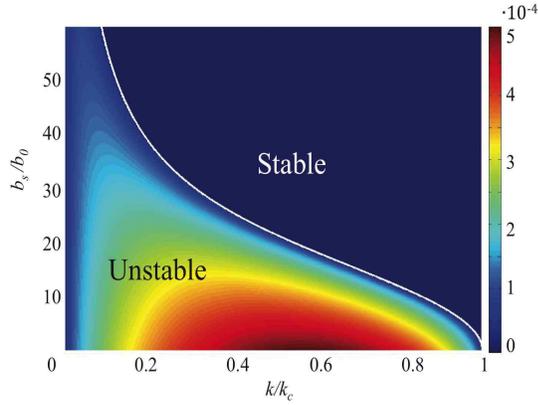}
\caption{Stability limits (white line) in the parameter space of the scaled wave number $k/{k_c}$  and the oscillation amplitude ${b_s}/{b_0}$, where ${b_0} = 2.5 \cdot {10^{ - 5}}$, ${\rm{\Omega }} = 20{\rm{max}}({\alpha _0})$. Shading indicates value of the instability growth rate in the unstable region, Eq. (\ref{eq49}).}
\end{figure}
\begin{figure}
\includegraphics[width=3.5in,height=2.4in]{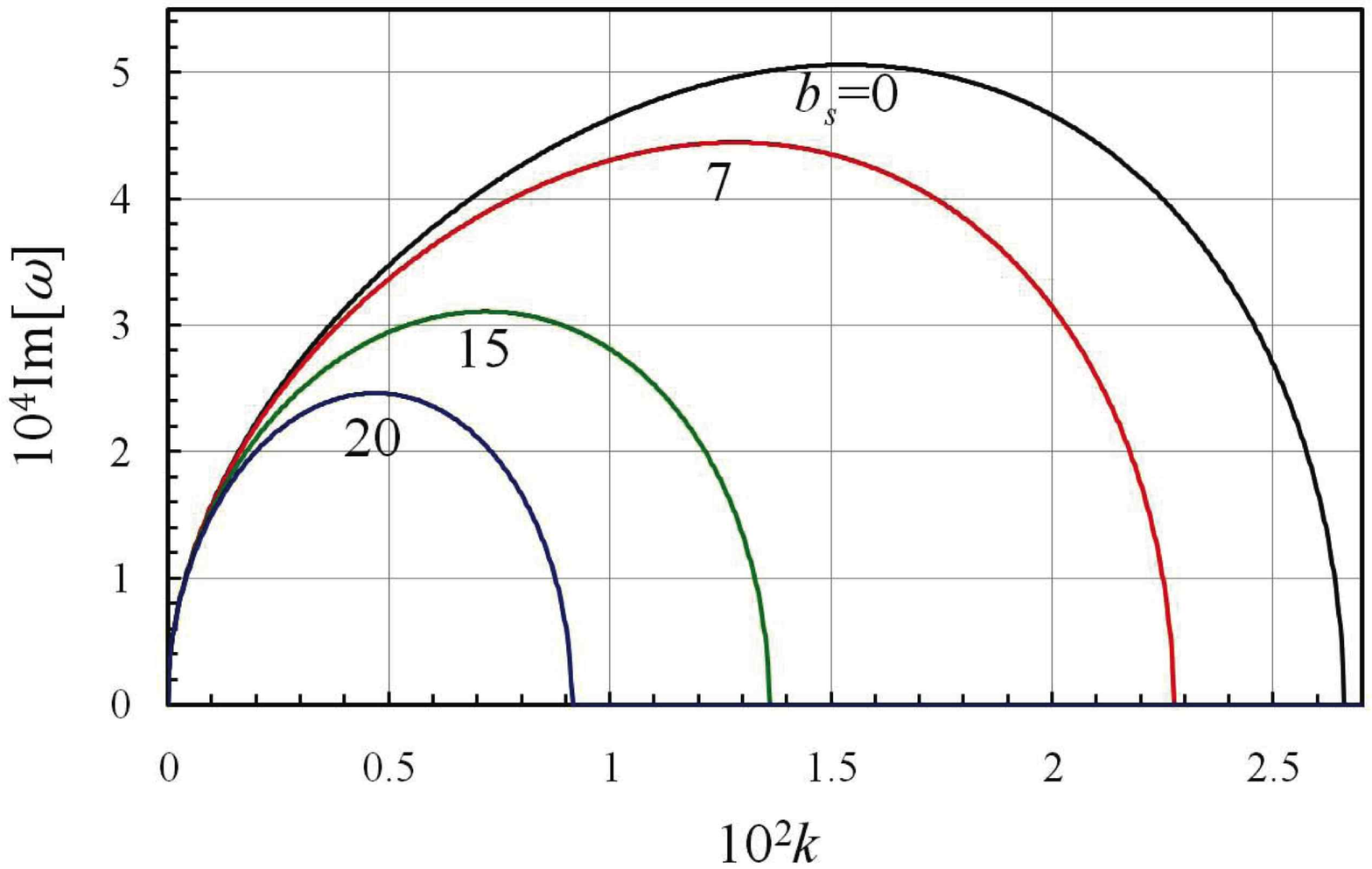}
\caption{The RT instability growth rate, Eq. (\ref{eq49}), in presence of external high-frequency field with ${\rm{\Omega }} = 20{\rm{max}}({\alpha _0})$ vs perturbation wave number for different values of ${b_1}/{b_0} = 0;7;15;20$ and ${b_0} = 2.5 \cdot {10^{ - 5}}$.}
\end{figure}

The second effect related to the oscillating external field is the parametric instability, which is known from mechanics \cite{bib-18} and which has been encountered in gas dynamics, e.g., of combustion systems \cite{bib-19,Searby1992,SearbyRochwerger1991}. The constant part of the "acceleration" is not required to obtain the parametric instability and it may be omitted in the calculations by taking ${b_0} = 0$. Similar to \cite{bib-18,bib-19} we look for a solution to Eq. (\ref{eq40}) in the form $\zeta  = \left[ {{\xi _1}\cos \left( {{\rm{\Omega }}t/2} \right) + {\xi _2}\sin \left( {{\rm{\Omega }}t/2} \right)} \right]\exp \left( {{\rm{\alpha }}t} \right)$. Mark that the growth of the parametric instability is accompanied by oscillations of the perturbation with frequency ${\rm{\Omega }}/2$, i.e. half of the driving force frequency. Substituting $\zeta $ into Eq. (\ref{eq40}) we obtain
\begin{eqnarray*}
\left( {{\alpha _0}^2 - {{{{\rm{\Omega }}^2}} \over 4} + {\rm{\Omega }}_{\rm{c}}^2} \right)\left[ {{\xi _1}\cos \left( {{{{\rm{\Omega }}t} \over 2}} \right) + {\xi _2}\sin \left( {{{{\rm{\Omega }}t} \over 2}} \right)} \right] \\
+ {\alpha _0}{\rm{\Omega }}\left[ { - {\xi _1}\sin \left( {{{{\rm{\Omega }}t} \over 2}} \right) + {\xi _2}\cos \left( {{{{\rm{\Omega }}t} \over 2}} \right)} \right] \\
= {b_s}k\cos \left( {{{{\rm{\Omega }}t} \over 2}} \right)\left[ {{\xi _1}\cos \left( {{{{\rm{\Omega }}t} \over 2}} \right) + {\xi _2}\sin \left( {{{{\rm{\Omega }}t} \over 2}} \right)} \right].
\end{eqnarray*}
Separating the terms with $\cos \left( {{\rm{\Omega }}t/2} \right)$ and $\sin \left( {{\rm{\Omega }}t/2} \right)$ and omitting higher-frequency terms, we reduce the equation to the linear algebraic system for the amplitudes ${\xi _1}$ and ${\xi _2}$
\begin{eqnarray*}
\left( {{\alpha _0}^2 - {{{{\rm{\Omega }}^2}} \over 4} + {\rm{\Omega }}_{\rm{c}}^2 - {{{b_s}k} \over 2}} \right){\xi _1} =  - \alpha {\rm{\Omega }}{\xi _2},\\
\left( {{\alpha _0}^2 - {{{{\rm{\Omega }}^2}} \over 4} + {\rm{\Omega }}_{\rm{c}}^2 + {{{b_s}k} \over 2}} \right){\xi _2} = \alpha {\rm{\Omega }}{\xi _1}.
\end{eqnarray*}
Solving the system we obtain the dispersion relation
\begin{equation}
\label{eq52}
{\left( {{\alpha ^2} - {{{{\rm{\Omega }}^2}} \over 4} + {\rm{\Omega }}_{\rm{c}}^2} \right)^2} - {{b_s^2{k^2}} \over 4} + {\alpha ^2}{{\rm{\Omega }}^2} = 0.
\end{equation}
Substituting $\alpha  = 0$ into Eq. (\ref{eq52}) we find the limiting oscillation amplitude ${b_s}$ required to induce the parametric instability
\begin{equation}
\label{eq53}
{b_s} > {2 \over k}\left| {{\rm{\Omega }}_{\rm{c}}^2 - {{{{\rm{\Omega }}^2}} \over 4}} \right|.
\end{equation}
In dimensional variables the last relation reads as
\begin{equation}
\label{eq54}
{{{\mu _B}B{'_s}} \over m} > {k^{ - 1}}\left| {4{\rm{\tilde \Omega }}_{\rm{c}}^2 - {{{\rm{\tilde \Omega }}}^2}} \right|,
\end{equation}
where the dimensional frequency of capillary waves is
$${{\rm{\tilde \Omega }}_c}\left( {\tilde k} \right) = {\tilde k^3}{{{\hbar ^2}\tilde n_0^{1/2}} \over {2{m^2}}}\sqrt {2\pi \left( {{a_{12}} - a} \right).} $$
\begin{figure}
\includegraphics[width=2.9in,height=2.4in]{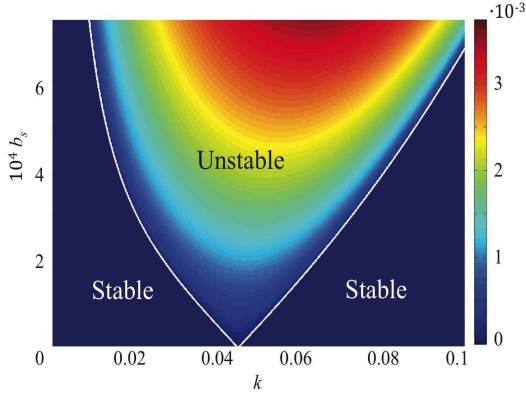}
\caption{Stability limits (white line) in the parameter space of the wave number $k$ and amplitude ${b_s}$ of the external high-frequency force with ${\rm{\Omega }} = 3.6 \cdot {10^{ - 3}}$. Shading indicates value of the instability growth rate in the unstable region, obtained from Eq. (\ref{eq52}).}
\end{figure}
\begin{figure}
\includegraphics[width=3.5in,height=2.0in]{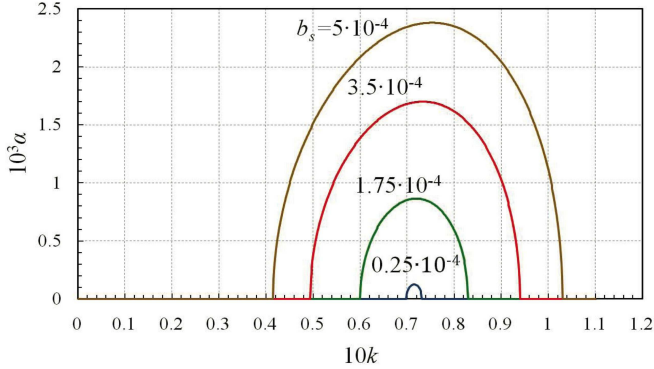}
\caption{Growth rate of the parametric instability, Eq. (\ref{eq52}), as function of $k$ for different amplitudes of the external oscillating magnetic force ${b_s} = 0.25 \cdot {10^{ - 4}};1.75 \cdot {10^{ - 4}};3.5 \cdot {10^{ - 4}};5 \cdot {10^{ - 4}}$ with ${\rm{\Omega }} = 3.6 \cdot {10^{ - 3}}$.}
\end{figure}
Minimal (infinitely small) critical amplitude of the magnetic field corresponds to the perturbations with the wave number determined by the equation ${{\rm{\Omega }}_c}(k) = {\rm{\Omega }}/2$. Therefore, when applying an external magnetic force $b\left( t \right) = {b_s}\cos \left( {{\rm{\Omega }}t} \right)$ to the system of two BECs, we should expect resonance pumping of the capillary waves with frequency ${\rm{\Omega }}/2$ and the wave number  $k = {{\rm{\Omega }}^{2/3}}{(2\gamma )^{1/6}}$. If capillary waves of another non-resonant wave number dominate initially, then pumping of the waves requires finite (not necessarily small) amplitude of the driving force. The instability domain is shown in Fig. 8 in the parameter space of the force amplitude and the wave number with the shading indicating the perturbation growth rate. Figure 9 presents the dispersion relation for the parametric instability (the growth rate versus the wave number) for different amplitude values of the oscillating magnetic field.

\subsection{RM instability produced by a magnetic pulse}
Within the context of quantum gases, the RM instability produced by a magnetic pulse $b\left( t \right) = {b_d}\delta \left( t \right)$ has been discussed recently in Ref. \cite{bib-14}, which indicated considerable differences in the instability development in BECs in comparison with the traditional configuration of classical gases separated by an inert interface. Here we analyze the linear interface dynamics by a more general (matrix) method of solving Eq. (\ref{eq42}) applicable to a wide class of time-dependent forces. Particularly, the matrix method will be extended to the case of stochastic forces in the next subsection. We introduce the vector function $\zeta_{v} = {\left( {\matrix{\zeta  & {\dot \zeta }\cr} } \right)^T}$ and rewrite Eq. (\ref{eq42}) as
\begin{equation}
\label{eq55}
{d \over {dt}}{\zeta_{v}}  = \left( {\matrix{  0 & 1  \cr { - {\rm{\Omega }}_c^2 + kb(t)} & 0  \cr } } \right)\zeta_{v},
\end{equation}
where the matrix may be split into a constant and time-dependent parts.
First, Eq. (\ref{eq55}) is solved for the case $b=0$, which yields
\begin{eqnarray}
\label{eq56}
{\zeta}_{v}(t)={\rm{exp}}\left[ {t\left( {\matrix{0 & 1  \cr  - \Omega_c^2 & 0\cr} } \right)} \right]{\zeta}_{v}(0) \nonumber \\
\nonumber \\
= \left( {\matrix{{\cos \left( {{{\rm{\Omega }}_c}t} \right)} & {\Omega_c^{-1}\sin \left( {{{\rm{\Omega }}_c}t} \right)}  \cr { - {{\rm{\Omega }}_c}\sin \left( {{{\rm{\Omega }}_c}t} \right)} &{\cos\left({{{\rm{\Omega}}_c}t}\right)}  \cr  } } \right){\zeta}_{v}(0).
\end{eqnarray}
For convenience, we introduce the designations
\begin{equation}
\label{eq-Vdefinition}
\textbf{V}_{0}\equiv \left({\matrix{0 & 0\cr1 & 0 \cr}}\right),
\end{equation}
\begin{equation}
\label{eq}
{{\bf{G}}_0}(t)\equiv \left( {\matrix{{\cos \left( {{{\rm{\Omega }}_c}t} \right)} & {\Omega_c^{-1}\sin \left( {{{\rm{\Omega }}_c}t} \right)}  \cr { - {{\rm{\Omega }}_c}\sin \left( {{{\rm{\Omega }}_c}t} \right)} &{\cos\left({{{\rm{\Omega}}_c}t}\right)}  \cr  } } \right).
\end{equation}
We look for a general solution to Eq. (\ref{eq55}) in the form $\zeta_{v} \left( t \right) = {{\bf{G}}_0}\left( t \right)\nu \left( t \right)$ and reduce Eq. (\ref{eq55}) to
\begin{equation}
\label{eq57}
\dot \nu \left( t \right) = kb(t){\mathbf{V}}(t)\nu(t),
\end{equation}
where
$${\mathbf{V}}(t) = {\bf{G}}_0^{ - 1}\left( t \right)\textbf{V}_{0}{{\bf{G}}_0}\left( t \right)$$
$$ = \left( {\matrix{   { - \left({2{{\rm{\Omega }}_c}}\right)^{-1}\sin \left( {2{{\rm{\Omega }}_c}t} \right)} & {- {{\rm{\Omega }}_c^{-2}}{\rm{si}}{{\rm{n}}^2}\left( {{{\rm{\Omega }}_c}t} \right)}  \cr   {{\rm{co}}{{\rm{s}}^2}\left( {{{\rm{\Omega }}_c}t} \right)} & { {(2{\rm{\Omega }}_c)}^{-1}\sin \left({2{{\rm{\Omega }}_c}t} \right)}  \cr } } \right).$$
The solution to Eq.(\ref{eq57}) is the time-ordered exponential ${\bf{F}}\left( \tau  \right)$, e.g. see \cite{bib-20},
\begin{eqnarray}
\label{eq58}
\exp \left[ { \int \limits_0^t d\tau {\bf{F}}(\tau )} \right] \equiv 1+ \nonumber \\
\sum \limits_{n = 1}^{ + \infty }  \int \limits_0^t d{t_1} \int \limits_0^{{t_1}} d{t_2} \cdots  \int \limits_0^{{t_{n - 1}}} d{t_n}{\bf{F}}({t_1}){\bf{F}}({t_2}) \cdots {\bf{F}}({t_n}).
\end{eqnarray}
We solve Eq. (\ref{eq57}), go back to the original variable and obtain the general solution to Eq. (\ref{eq55})
\begin{equation}
\label{eq59}
{\mathbf{\zeta}}_{v} \left( t \right) = {{\bf{G}}_0}\left( t \right)\exp \left\{ { \int \limits_0^t ds\,kb(s){\mathbf{V}}(s)} \right\}\mathbf{\zeta}_{v} (0).
\end{equation}
On the basis of the general solution, we may recover the results of Ref. \cite{bib-14} on the RM instability triggered by a pulse $b( t ) = {b_d}\delta ( t )$. In this case all terms of time-ordered exponential commute with each other and we obtain
$$\exp \left\{ { \int \limits_0^t ds\,k{b_d}\delta(s){\mathbf{V}}(s)} \right\}$$
$$ = \exp \left( {\matrix{   0 & 0  \cr   {k{b_d}} & 0  \cr } } \right) = \left( {\matrix{   1 & 0  \cr    {k{b_d}}& 1  \cr } } \right),$$
which yields the final solution
\begin{equation}
\label{eq60}
\zeta(t) = \sqrt {{\zeta ^2}\left( 0 \right) + {1 \over {{\rm{\Omega }}_c^2}}{{\left[ {k{b_d}\zeta \left( 0 \right) + \dot \zeta \left( 0 \right)} \right]}^2}} \sin \left( {{{\rm{\Omega }}_c}t + {\varphi _d}} \right)
\end{equation}
with the new oscillation phase determined by
\begin{equation}
\label{eq61}
{\rm{tan}}\,{\varphi _d} = {{{{\rm{\Omega }}_c}\zeta \left( 0 \right)} \over {\dot \zeta \left( 0 \right) + k{b_d}\zeta \left( 0 \right)}}.
\end{equation}
Thus, the general solution Eq. (\ref{eq59}) reproduces the analytical results of Ref. \cite{bib-14} given by Eq. (\ref{eq60}). Particularly, Eq. (\ref{eq60}) describes the jump in the amplitude and phase of a standing capillary wave because of the pulse for a non-zero bending of the interface,  $\zeta \left( 0 \right)\neq 0$. In the opposite case of  $\zeta \left( 0 \right)= 0$ and  $\dot\zeta \left( 0 \right)\neq 0$ the pulse does not influence the interface dynamics. Eq. (\ref{eq60}) shows that the transfer of magnetic energy of the pulse to superfluid oscillations is the most efficient when the deviation of the interface just before the pulse is maximal.
\begin{figure}
\includegraphics[width=3.4in,keepaspectratio]{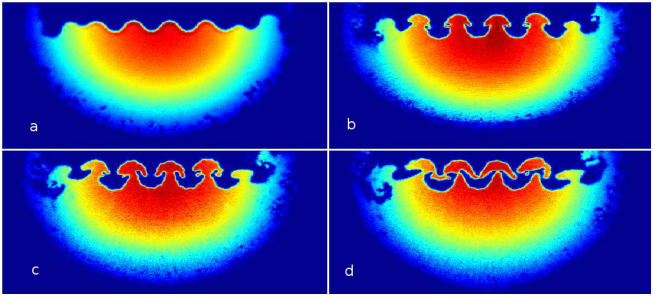}
\caption{Development of the RM instability in a trapped system of two-component BEC presented for one of the components.}
\end{figure}

The linear analysis works rigorously as long as the perturbation amplitude after the pulse remains sufficiently small. Once it becomes large enough, nonlinear effects dominate, and the interface evolves into a mushroom pattern with detachment of droplets as discussed in \cite{bib-14}. Figure 10 illustrates the development of the RM instability in BECs at the nonlinear stage obtained using the numerical methods similar to \cite{bib-14}. We use the same system of BECs as in Fig. 4. A magnetic pulse of duration $\tau  = 0.02{\rm{ms}}$ and size $\sqrt \pi  {\mu _B}B'\tau /2m = 1.05 \cdot {10^{ - 3}}{\rm{m}}/{\rm{s}}$ is applied to the initial capillary wave seen in Fig. 10 (a). The following graphs correspond to the time instants 2.39; 4.77; and 7.16 ms after the pulse (figures b, c, d, respectively). Development of the RM instability at the initial stages (e.g. Fig. 10 b) resembles the RT instability shown in Fig. 4. However, the difference between two instabilities is noticeable at later stages shown in Fig. 10 (c, d). In particular, we observe the detachment of droplets in the RM instability in agreement with the model presented in Ref.\ \cite{bib-14}.

\subsection{Gaussian noise}
Finally, we consider the interface dynamics under the action of a Gaussian random force ${b_r}\left( t \right)$ with mean zero $\left\langle {b_r}\left( t \right)\right\rangle = 0$, where the brackets imply averaging over a statistical ensemble of realizations.
The force is switched on at $t = 0$. Because of the Gaussian property, the force is characterized only by variance $Q$ and by the correlation time ${\tau _c}$. The case of zero ${\tau _c}$ is known as the Markovian case, and in this case averaging may be performed exactly. On the contrary, the case of non-zero ${\tau _c}$ is not solvable in general, but an analytical approximate solution may be obtained in the limit of small correlation time ${\tau _c}{{\rm{\Omega }}_{\rm{c}}} \ll 1$.

In order to obtain an averaged form of Eq. (\ref{eq40}) we employ the general solution (\ref{eq59}). Since Eq. (\ref{eq59}) is a time-ordered exponential with non-commutative kernel at different moments of time, we use the cumulant method  for averaging \cite{bib-20}. Within this method, if $\textbf{G}(\tau )$ is a stochastic Gaussian operator with mean zero $\left\langle\textbf{G}(t)\right\rangle = 0$, then the time-ordered exponential of $\textbf{G}(\tau )$ is averaged by the formula
\begin{equation}
\label{eq62}
\left\langle{\exp \left[ { \int \limits_0^t d\tau \mathbf{G}(\tau )} \right]}\right\rangle= \exp \left[ { \int \limits_0^t d\tau  \int \limits_0^\tau  ds \left\langle{\mathbf{G}(\tau )\mathbf{G}(s)}\right\rangle}\right].
\end{equation}
In the Markovian case this formula is exact, but in the non-Markovian case this is an approximation valid when ${\tau _c}$ is small compared to the intrinsic time scale of the system, ${\tau _c}{{\rm{\Omega }}_{\rm{c}}} \ll 1$, and the noise force is sufficiently weak to be treated as a perturbation \cite{bib-20}. The kernel of the right-hand side of Eq.(\ref{eq62}) is the cumulant series terminated at the first non-vanishing term (the first-order term vanishes because $\left\langle\textbf{G}(t)\right\rangle = 0$). We note that all cumulants in Eq. (\ref{eq62}) vanish identically for ${\tau _c} = 0$ because of the Gaussian property, but they are non-zero for finite ${\tau _c}$ due to non-commutativity of the kernel in Eq. (\ref{eq59}).

We start with the Markovian case when the autocorrelation function is given by the delta-function
\begin{equation}
\label{eq63}
\left\langle {{b_r}(t){b_r}(s)} \right\rangle  = Q\,\delta (t - s).
\end{equation}
Using Eqs. (\ref{eq59}), (\ref{eq62}), we find that the averaged solution to Eq. (\ref{eq40}) reads
\begin{equation}
\label{eq64}
\left\langle \zeta ( t )\right\rangle = \zeta ( 0)\cos \left( {{{\rm{\Omega }}_c}t} \right) + {{\dot \zeta ( 0 )} \over {{{\rm{\Omega }}_c}}}\sin \left( {{{\rm{\Omega }}_c}t} \right),
\end{equation}
\emph{i.e.} the Markovian stochastic force does not influence averaged dynamics of the interface.

Now we consider a non-Markovian exponentially correlated noise with autocorrelation function given by
\begin{equation}
\label{eq65}
\left\langle {{b_r}( t ){b_r}( s )} \right\rangle  = {Q \over {2{\tau _c}}}{\rm{exp}}\left( { - \left| {t - s} \right|/{\tau _c}} \right),
\end{equation}
which reduces to the Markovian case in the limit of zero correlation time ${\tau _c} \to 0$. We average Eq. (\ref{eq40}) using the Novikov theorem \cite{bib-21}:
\begin{equation}
\label{eq66}
{{{d^2}} \over {d{t^2}}}\langle\zeta(t)\rangle  + {\Omega _c^2}\langle\zeta(t)\rangle  + k \int \limits_0^t ds\left\langle {{b_r}( t ){b_r}( s )} \right\rangle \left\langle{{\delta \zeta (t)} \over {\delta {{\tilde b}_r}( s )}}\right\rangle = 0.
\end{equation}
We calculate the functional derivative in Eq.(\ref{eq66}) using the general solution Eq.(\ref{eq59})
\begin{eqnarray}
&&{{\delta \mathbf{\zeta}_{v} (t)} \over {\delta {b_r}\left( s \right)}} = {\delta  \over {\delta {b_r}\left( s \right)}}{{\bf{G}}_0}\left( t \right)\exp \left\{ { \int \limits_0^t ds\,kb(s)\textbf{V}(s)} \right\}\zeta_{v} \left( 0 \right) \nonumber \\
&&= \theta(t-s)\theta(s){{\bf{G}}_0}(t){\bf{G}}_0^{-1}(s)k\textbf{V}_{0}{{\bf{G}}_0}( s )\nonumber \\
&&\times \exp \left\{ { \int \limits_0^t ds\,kb(s)\textbf{V}(s)} \right\}\zeta_{v} ( 0 ) + {\Delta }\left[ {{b_r}} \right]( {t,s} )\mathbf{\zeta}_{v} ( 0 )\nonumber\\
&&=\theta(t-s)\theta(s){{\bf{G}}_0}(t-s)k\textbf{V}_{0}{\bf{G}}_0^{ - 1}( {t - s} )\zeta_{v} ( t )\nonumber\\
&&+ {\rm{\Delta }}\left[ {{b_r}} \right]\left( {t,s} \right)\zeta_{v} ( 0),
\label{eq67}
\end{eqnarray}
where $\theta(t)$ is the Heaviside step function, and ${\Delta }\left[ {{b_r}} \right]\left( {t,s} \right)$ is a functional arising from the non-commutativity of the kernel of the exponential in Eq.(\ref{eq67}). The functional ${\Delta }\left[ {{b_r}} \right]\left( {t,s} \right)$ may be dropped in the present limit of weak noise. We obtain from Eq. (\ref{eq67})
\begin{eqnarray}
\left\langle {{{\delta \zeta (t)} \over {\delta {b_r}\left( s \right)}}} \right\rangle & \simeq &\theta(t - s)\theta(s)\,k\left\{ {\left[ \textbf{V}(t-s)\right]_{11} }\left\langle {\zeta (t)} \right\rangle +\right. \nonumber\\
&&\left.\left[\textbf{V}(t-s)\right]_{12}{d \over {dt}}\left\langle {\zeta ( t )} \right\rangle\right\}.
\label{eq68}
\end{eqnarray}
Substituting the operators ${{\bf{G}}_0}$ and $\textbf{V}_{0}$ to Eq. (\ref{eq68}), and using Eq. (\ref{eq66}) we find
\begin{eqnarray}
&&{{{d^2}} \over {d{t^2}}}\left\langle {\zeta (t)} \right\rangle  + {\Omega_c^2}\left\langle {\zeta (t)} \right\rangle+\nonumber\\
&&\int \limits_0^t ds{{{k^2}} \over {2{{\rm{\Omega }}_c}}}\left\langle {{b_r}( t){b_r}( s)} \right\rangle \sin 2{{\rm{\Omega }}_c}({t - s})\left\langle {\zeta \left( t \right)} \right\rangle-\nonumber\\
&&\int \limits_0^t ds{{{k^2}} \over {{\rm{\Omega }}_c^2}}\left\langle {{b_r}( t){b_r}( s)} \right\rangle {\sin ^2}{{\rm{\Omega }}_c}({t - s}){d \over {dt}}\left\langle {\zeta ( t )} \right\rangle  = 0.\nonumber\\
\label{eq69}
\end{eqnarray}
Calculating the integrals on time scales much larger than the correlation time of the noise  $t \gg {\tau _c}$, we arrive to an equation describing the averaged interface dynamics
\begin{equation}
\label{eq70}
{d^2 \over {dt^2}}\left\langle {\zeta ( t )} \right\rangle  - 2\kappa {d \over {dt}}\left\langle {\zeta ( t )} \right\rangle + \left( {{\rm{\Omega }}_c^2 + \nu _0^2} \right)\left\langle {\zeta (t)} \right\rangle  = 0,
\end{equation}
where
\begin{equation}
\label{eq71}
\kappa  = {1 \over 2}{{{k^2}Q\tau _c^2} \over {1 + {{\left( {2{{\rm{\Omega }}_c}{\tau _c}} \right)}^2}}}
\end{equation}
is the average growth rate and
\begin{equation}
\label{eq72}
\nu _0^2 = {1 \over 2}{{{k^2}Q{\tau _c}} \over {1 + {{\left( {2{{\rm{\Omega }}_c}{\tau _c}} \right)}^2}}}
\end{equation}
modifies the frequency  due to noise. The solution to Eq. (\ref{eq70}) grows exponentially in time as
\begin{equation}
\label{eq73}
\zeta  \propto \exp \left[ {t\left( {\kappa  \pm i\sqrt {{\rm{\Omega }}_c^2 + \nu _0^2 - {\kappa ^2}} } \right)} \right].
\end{equation}
Thus, a stochastic non-Markovian force leads to unstable growth of the average perturbation amplitude with modified frequency of the capillary oscillations.
\begin{figure}
\includegraphics[width=3.5in,height=2.0in]{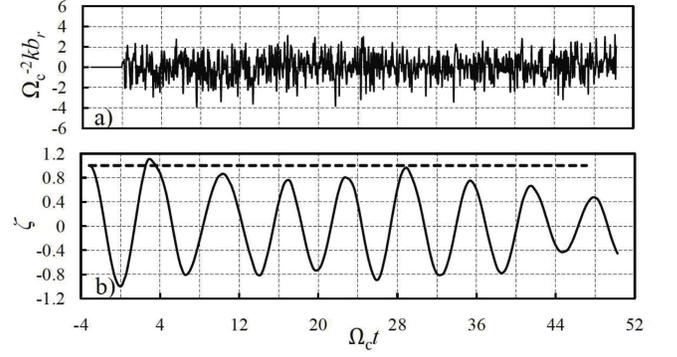}
\caption{(a) Single realization of the external stochastic Gaussian Markovian force with ${\rm{\Omega }} = 5 \cdot {10^{ - 9}}$, ${\tau _c} = 50$ and ${{\rm{\Omega }}_c}{\tau _c} = 1.5 \cdot {10^{ - 2}}$. (b) The respective numerical solution to Eq. (\ref{eq40}). The dashed line shows amplitude of the averaged analytical solution, Eq. (\ref{eq73}).}
\end{figure}
\begin{figure}
\includegraphics[width=3.5in,height=2.0in]{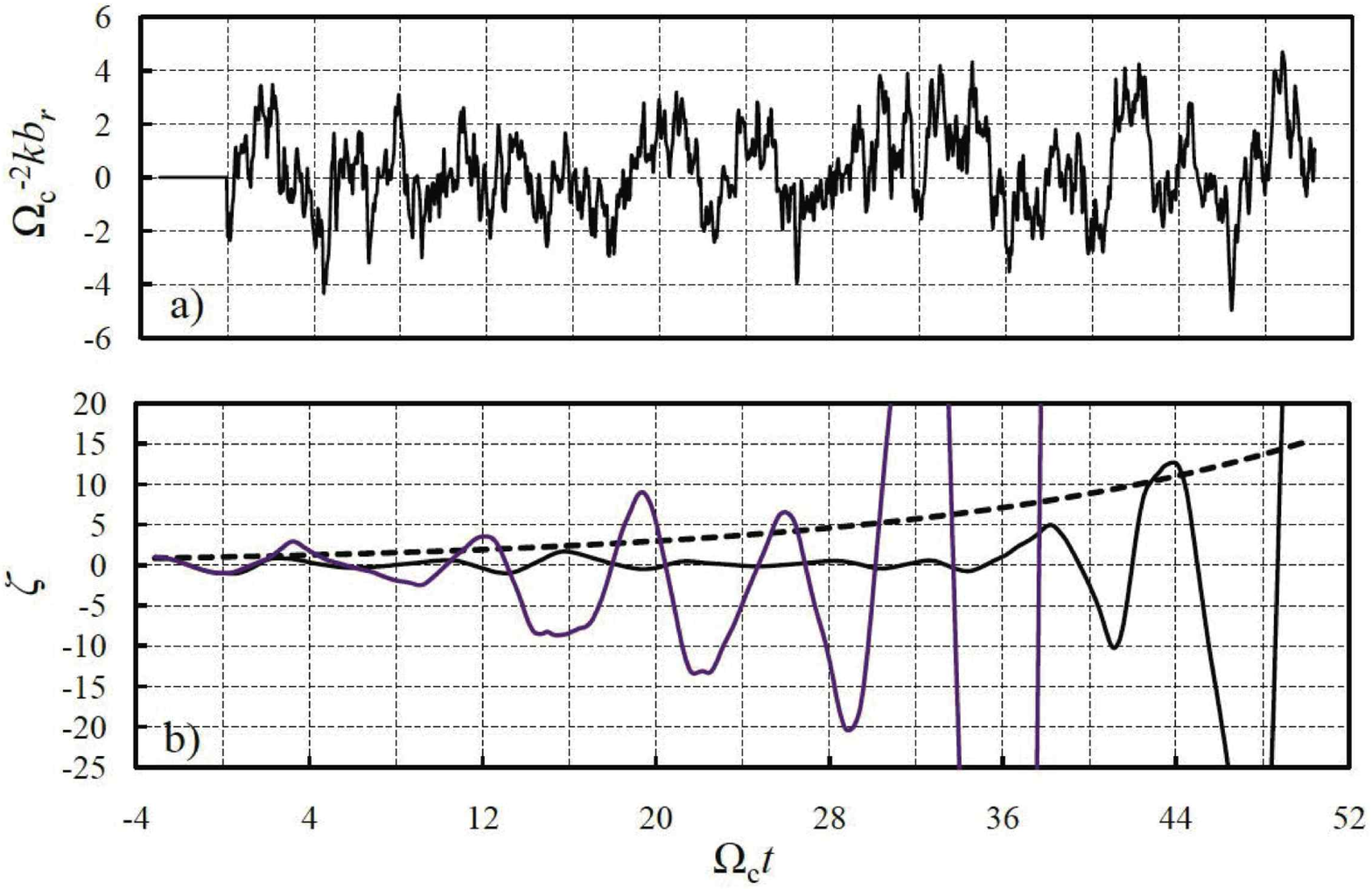}
\caption{(a) Single realization of the external stochastic Gaussian Markovian force with ${\rm{\Omega }} = 2 \cdot {10^{ - 7}}$ and ${\tau _c} = 1200$ and ${{\rm{\Omega }}_c}{\tau _c} = 0.38$. (b) The respective numerical solution to Eq. (\ref{eq40}) (two realizations). The dashed line shows amplitude of the averaged analytical solution, Eq. (\ref{eq73}).}
\end{figure}
The analytical theory for interface dynamics under the action of a stochastic force is compared in Figs. 11, 12 to the numerical solution to Eq. (\ref{eq40}) calculated for $k = 0.0154$. We note that real stochastic processes always have finite frequency band width, though it may be very large. Therefore, any real stochastic process is characterized by some non-zero correlation time  leading to exponential growth of amplitude of the oscillator Eq.(\ref{eq40}), which means that the Markovian case may be recovered only as a limit of ${{\rm{\Omega }}_c}{\tau _c} \to 0$. Figure 11 shows the numerical solution to Eq. (\ref{eq40}) for the stochastic force with $Q = 5 \cdot {10^{ - 9}}$ and ${{\rm{\Omega }}_c}{\tau _c} = 1.5 \cdot {10^{ - 2}} \ll 1$, which may be considered as a good approximation for the Markovian case. The dashed line represents the envelope function of the averaged amplitude of the interface oscillations predicted by the theory, Eqs. (\ref{eq71}), (\ref{eq73}). In the Markovian case, the averaged amplitude does not change, which is supported by the numerical results shown in Fig. 11. Figure 12 corresponds to a non-Markovian case with $Q = 2 \cdot {10^{ - 7}}$, ${\tau _c} = 1200$ and ${{\rm{\Omega }}_c}{\tau _c} = 0.38$. The numerical solution to Eq. (\ref{eq40}) is shown by the solid lines for two realizations of the noise, while the dashed line represents envelope of the averaged amplitude increasing exponentially in time according to the analytical theory Eqs. (\ref{eq71}), (\ref{eq73}). In agreement with the theoretical predictions, the numerical solution also demonstrates exponential increase of the perturbation amplitude.

\section{SUMMARY}

We have studied the dynamics of an interface in a two-component BEC driven by a spatially uniform time-dependent force. Experimentally, such a force may be produced by a magnetic field gradient acting on the system of two interacting BECs with different spins \cite{bib-10}. By applying the variational principle to the GP Lagrangian, we have derived the dispersion relation (or the respective ordinary differential equations) for linear waves and instabilities at the interface. We have considered different time-dependent forces leading to a diverse collection of dynamical effects:
(i)	A constant force pushing the components towards each other generates the RT instability for a certain domain of perturbation wave numbers. Because of the RT instability, small perturbations grow exponentially in time without oscillations.
(ii)	A sinusoidal force with frequency ${\rm{\Omega }}$ leads to the parametric instability at the interface. The critical strength of the force required to drive the instability depends on the perturbation wave number. This strength is infinitely small for the intrinsic capillary waves at the interface with frequency equal to ${\rm{\Omega }}/2$. Because of the parametric instability, small perturbations not only grow exponentially in time, but also oscillate with frequency ${\rm{\Omega }}/2$.
(iii)	A pulse force leads to the quantum counterpart of the RM instability. Within the linear approximation, the force produces a discontinuous jump in the amplitude and phase of the capillary waves at the interface.
(iv) A non-Markovian stochastic external field (with non-zero correlation time) drives the instability at the interface accompanied by oscillations. The growth rate of the instability is determined by the variance of the driving force and by the correlation time. The oscillation frequency is shifted in comparison to the intrinsic frequency of the capillary waves. A Markovian force on average does not lead to any effect.

\acknowledgements

This research was supported partly by the Swedish Research Council (VR) and by the Kempe Foundation. Calculations have been conducted using the resources of High Performance Computing Center North (HPC2N).


\end{document}